\documentclass[twocolumn,aps,pra,groupedaddress,showpacs]{revtex4}
\usepackage{epsfig,amssymb,amsmath}
\begin{document}    

\title{Inequalities for quantum channels assisted by
limited resources}

\author{Vittorio Giovannetti}
\affiliation{NEST-INFM \& Scuola Normale Superiore,
piazza dei
Cavalieri 7, I-56126 Pisa, Italy.}

\begin{abstract}
The information capacities and ``distillability''
of a quantum channel are studied
in the presence of auxiliary resources. These include prior 
entanglement shared
between
the sender and receiver and free classical bits of forward and backward 
communication. Inequalities and trade-off curves are derived. In particular
an alternative proof is given
 that in the absence of
feedback and shared entanglement,
forward classical communication does not increase the quantum
capacity of a channel. 
\end{abstract}
\maketitle 

Any realistic scheme for information transmission
must take into account the presence of noise. 
Given the technological challenge we are facing in controlling 
decoherence (e.g.~\cite{CHUANG} and references therein)
this is even more important when
transmitting quantum information through quantum
channels~\cite{SHOR}. 
One way to reduce the effects of noise is to provide
the communicating parties with some extra resource that
can be used to implement more efficient communication
protocols.
It is known, for instance, that  
teleportation~\cite{TELE} and superdense coding~\cite{SUP}
can increase both
the quantum and classical capacities of a channel 
by allowing the sender and receiver
of the message to share a sufficient amount of prior
entanglement~\cite{BENNETT1,SHORCE}.
Alternatively, using entanglement distillation protocols~\cite{DIST,DIST1}, 
the channel performances
can be improved by introducing  a classical feedback side channel~\cite{BOWEN} 
or by allowing  the
sender and receiver to communicate freely through a classical 
two-way side channel ~\cite{SHOR,BENNETT1}.
Following the suggestion of Refs.~\cite{DEVETAK,SHORCE,DHW,HARROW}
in this paper we analyze the relationships between 
different resources by  focusing on the case where resources are limited.

The material is organized as follows. In Sec.~\ref{s:def} 
we introduce the
notation, define the distillability
of a channel in the presence of finite resources, 
and establish some preliminary results. In Sec.~\ref{s:quno}
we give a new proof that, in the absence of prior entanglement and
feedback,
free forward classical  communication does not 
increase the quantum capacity of a channel. 
In Sec.~\ref{s:tradeoff} we study
the quantum and classical capacities as functions of the 
resource parameters and we establish some trade-offs and asymptotic 
limits. In Sec.~\ref{s:disto} we provide some identities
for the distillability: in particular we show
that with only free classical forward communication
the distillability of a channel cannot be greater 
than its unassisted quantum capacity.
The paper ends in Sec.~\ref{s:con} with the conclusion.

\section{Quantum channels with limited resources}\label{s:def}

Consider a memoryless 
quantum channel~\cite{SHOR} 
described by a Completely Positive, Trace preserving
(CPT) map $\cal M$ defined in a $d$-dimensional Hilbert space $\cal H$. 
For the sake of simplicity we will measure the capacities 
of such channel in ``dits'' or ``qudits'' per channel use, 
where $1$ dit stands for
$\log_2 d$ bits of classical information and $1$ qudits
for $\log_2 d$ qubits of quantum information. 
Analogously we define $1$ ``e-dit''
the entanglement associated with a maximally entangled
state of ${\cal H} \otimes {\cal H}$, i.e. $\log_2 d$ e-bits.
\begin{figure}[t]
\begin{center}
\epsfxsize=.80\hsize\leavevmode\epsffile{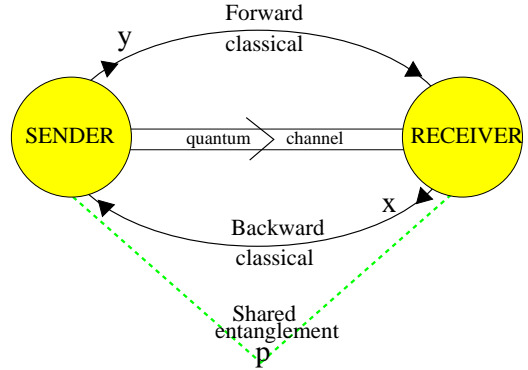}
\end{center}
\caption{Scheme of the communication scenario:
the sender transmits quantum or classical information
to the receiver through the quantum channel $\cal M$. 
On average, for each channel
use they are allowed to employ $x \log_2 d$ bits of classical feedback,
$y \log_2 d$ bits of classical forward information, and $p \log_2d$ e-bits
of shared entanglement. With the help of these extra resources, the two
communicating parties can increase the capacities of the channel by using
purification protocols~\cite{DIST,DIST1}, teleportation~\cite{TELE},
 superdense coding~\cite{SUP}, etc.
 }
\label{f:fig1}
\end{figure}

We are interested in the scenario depicted in Fig.~\ref{f:fig1}. 
At each channel use
the sender
and the receiver are provided on {\em average}  with  $x$ dits of 
backward (i.e. from the receiver to the sender) 
classical communication, $y$ dits of forward (i.e. from the
sender to the receiver) classical communication,
and $p$ e-dits of shared entanglement.
A rigorous definition of $x$, $y$ and $p$ requires one
to consider a limit $N\rightarrow \infty$ in the total number $N$
of channel uses. That is, if $X_N$ is the total number of dits
of classical feedback available on $N$ uses of $\cal M$, we define
$x$ as $\lim_{N\rightarrow \infty} X_N/N$. Analogous definitions
apply for $y$ and $p$.
Moreover in defining $y$ we do not assume the
classical information transmitted through the side channel 
to be independent of
the classical message transmitted through $\cal M$. 
This is different from the definition adopted 
in~\cite{BENNEW} where the communicating parties cannot use the side
channel to directly transfer part of the message (in their case however $y$ is
ideally infinite and our definition would produce only trivial results).

Exploiting the resources $x$, $y$ and $p$  
the two communicating parties can improve
the performances of the channel $\cal M$.
In particular by means of $x$ and $y$ they can
set up purification protocols to augment the number $p$ of
shared maximally entangled states.
On the other hand they  can employ $p$ in
teleportation and superdense coding schemes to transfer
reliably (i.e. with unit fidelity)
quantum or classical information from the sender to
the receiver. 
Alternatively, $p$ and $x$ can be used to create a 
quantum feedback connection through teleportation.
In general the optimal
strategy associated with the resources $x$, $y$ and $p$ 
consists in some complicated composition of all these effects.

We  call $Q(x,y,p)$
the quantum capacity achievable
in the communication scenario of 
Fig.~\ref{f:fig1}, while we use $C(x,y,p)$ to indicate
the corresponding classical capacity.
These two objects
represent respectively the maximum number of {\em unknown} 
qudits or dits 
that can be reliably transmitted through
$\cal M$ per channel use~\cite{SHOR}.
We also introduce the 
``distillability'' $D(x,y,p)$ which
gives, in e-dits per channel use, the maximum number
of  maximally entangled states
of ${\cal H}\otimes {\cal H}$ 
that can be asymptotically 
shared between the sender and receiver 
by using $\cal M$ 
and the resources $x$, $y$ and $p$.
Such a quantity is not a proper capacity of $\cal M$,
as the communicating parties know {\em a priori} which states  
(i.e. the maximally entangled states)
they are going to share.
Apart from regularization over multiple uses 
(see Appendix~\ref{s:app2} for details) $D(x,y,p)$ 
can be expressed as 
\begin{eqnarray}
D(x,y,p) &=& \lim_{n\rightarrow \infty} \max_{R} \big\{
\frac{P_{x,y,p}\left[({\cal M}\otimes\openone)^{\otimes n} 
(R)\right]}{n} \big\}\;,
\label{distillato}
\end{eqnarray}
where the maximization is performed over all density matrices $R$ defined 
in the Hilbert space ${\cal H}^{\otimes n}\otimes {\cal H}^{\otimes n}$.
In this equation  $R^\prime\equiv 
({\cal M}\otimes\openone)^{\otimes n} (R)$ is
the state we get by sending half of $R$ through $n$ copies of 
the channel $\cal M$ and
doing nothing (i.e. applying the 
identity superoperator $\openone$) to the other half.
Finally the quantity
$P_{x,y,p}\left[R^\prime\right]$
is the maximum number of e-dits that can be 
asymptotically 
extracted from $R^\prime$ through purification protocols
which employs, on average,  $x$ dits of feedback, $y$ dits of classical
forward communication and $p$ e-dits of prior shared entanglement
for any use of  the channel $\cal M$.
In other words, $D(x,y,p)$ is obtained by maximizing over all possible
input $R$ the distillability of the output state $R'$ achievable
by using protocols that employ $x$, $y$ and $p$ resources.

The aim of this paper is to study the 
dependence of $Q(x,y,p)$, $C(x,y,p)$ and
$D(x,y,p)$ upon
the variables $x$, $y$ and $p$.
Such an endeavor is connected with the study of the channel 
capacity for the simultaneous transmission of quantum and
classical information~\cite{DEVETAK} since, for instance,
we can interpret the resource $y$ as
the classical information transmitted through $\cal M$ in some
previous channel use.

\subsection{Basic properties}\label{s:basic}
The classical capacity and distillability provide two
trivial upper
 bounds for the quantum capacity 
of the channel, i.e.
\begin{eqnarray}
C(x,y,p) &\geqslant& Q(x,y,p)\;,\nonumber\\
D(x,y,p) &\geqslant& Q(x,y,p)\;.
\label{distillato1}
\end{eqnarray}
In fact, on the  one hand, at each channel use
 we can transmit  $Q(x,y,p)$ dits of classical information
by encoding them into qudits. On the other hand,
at each channel use we can produce  $Q(x,y,p)$ e-bits 
between the sender and the receiver by
transmitting $Q(x,y,p)$ halves of maximally entangled states
of ${\cal H}\otimes {\cal H}$.
The
relation between $D(x,y,p)$ and $C(x,y,p)$ is more complex and
even though there
are situations in which the later is bigger than the former, we are
not able to provide a definitive ordering
 (see also Sec.~\ref{s:disto}).

The quantities $Q(x,y,p)$, $C(x,y,p)$ and $D(x,y,p)$ 
are non-decreasing,
jointly-concave functions of their arguments.
The first property derives simply 
from the fact that the sender and receiver 
can discard part of the
resources they are given to exactly simulate communication scenarios with
fewer initial resources. The concavity derives instead 
from the possibility that the communicating
parties use their resources within time-sharing strategies
(see App.~\ref{s:app1}).

\paragraph*{Unassisted capacities:--} For $x,y,p=0$  
the capacities defined above give
the unassisted capacities of $\cal M$, i.e.
$Q\equiv Q(0,0,0)$ and $C\equiv C(0,0,0)$.
These quantities can be determined by maximizing (over
multiple channel uses) 
the coherent information~\cite{SETHQ}  
and the output Holevo information of the channel
\cite{HSW}, respectively. 
\paragraph*{Entanglement assisted capacities:--}
The functions $Q(0,0,p)$ and $C(0,0,p)$ represent 
the entanglement assisted capacities of $\cal M$ in the absence of
classical feedback and classical forward communication. 
Shor recently gave~\cite{SHORCE} a procedure to compute the value
of $C(0,0,p)$ while a method to calculate $Q(0,0,p)$ is provided by
Devetak, Harrow and Winter in Ref.~\cite{DHW}.
In the limit of large $p$ it has been shown~\cite{BENNETT1} that
$C_E\equiv C(0,0,\infty)$
can be obtained by maximizing the quantum mutual
information~\cite{ADAMI} of the channel while,  
due to teleportation and superdense coding,
\begin{eqnarray}
Q_E = C_E/2\;.
\label{utile}
\end{eqnarray} 
The minimum values  ${\cal E}_Q$ and ${\cal E}_C$  of $p$
for which  $Q(0,0,p)$ and $C(0,0,p)$ 
achieve respectively $Q_E$ and $C_E$ are known to be 
of the order of one e-dit per channel use~\mbox{\cite{SHORCE,BENNETT1}}.
Moreover  the following relations have been established~\cite{BOWEN1}
\begin{eqnarray}
 {\cal E}_C +  Q_E  \geqslant  {\cal E}_Q \geqslant {\cal E}_C - Q_E 
\label{bowenin1}\;, \\
 {\cal E}_Q  \geqslant  Q_E - Q\;, \quad 
  {\cal E}_C \geqslant  C_E - C \label{bowenin2}\;.
\end{eqnarray}
\paragraph*{Feedback and entanglement:--}
In the limit of large $p$, Bowen~\cite{BOWEN} 
studied the effect of an arbitrary amount of
classical feedback $x$ on the quantum and classical 
capacity of a channel.
These results can be summarized in our formalism
by the following relations,
\begin{eqnarray}
C(x,0,p) &=&  C(0,0,\infty) \equiv C_E \quad
\mbox{for}\; p \geqslant{\cal E}_C   
\nonumber \\
Q(x,0,p) &=&  Q(0,0,\infty) \equiv Q_E \quad
\mbox{for} \; p \geqslant{\cal E}_Q \;,
\label{bow1}
\end{eqnarray}
which imply that, in the absence of
forward classical communication $(y=0)$, feedback cannot be
used to increase the capacities above the level achieved 
with arbitrary shared entanglement.
\paragraph*{Feedback and forward communication:--}  
For $p=0$ and 
arbitrary backward and forward classical communication
$Q(x,y,p)$ gives the two-way quantum capacity,
$Q_{2\mbox{\tiny{-way}}} \equiv Q(\infty,\infty,0)$
\cite{BENNETT1,SHOR,DIST}. There is no simple recipe
 to compute $Q_{2\mbox{\tiny{-way}}}$, but
using teleportation one can show that 
 this capacity coincides with the maximum amount of
maximally entangled state
that can be shared between
the sender and receiver
using arbitrary two-way purification protocols, i.e.
\begin{eqnarray}
Q(\infty,\infty,0) = D(\infty,\infty,0) \;,
\label{qudue}
\end{eqnarray}
with $D(\infty,\infty,0)$ the distillability  
of Eq.~(\ref{distillato})
evaluated for $x=y=\infty$ and $p=0$.

The relations between the quantities defined above have not yet completely
understood. In the case of quantum capacities, we know 
for instance that $Q_E$ and $Q_{2{\mbox{\tiny{-way}}}}$
are always greater than or equal to $Q$ and recently it has been proved
\cite{BENNEW}
that $Q_E \geqslant Q_{2\mbox{\tiny{-way}}}$.
Equation~(\ref{bow1}) establishes that  $Q_E$ is
greater than or equal to 
the quantum capacity of the channel in the presence of
arbitrary amount of feedback $Q_{FB}\equiv Q(\infty,0,0)$, 
but it is unclear~\cite{BOWEN,BENNEW} if this last quantity is 
strictly smaller than $Q_{2{\mbox{\tiny{-way}}}}$.

In the following sections we will try to characterize the relations
among communication scenarios assisted by different initial resources. 
We begin in Sec.~\ref{s:quno} by deriving two simple identities which
involve forward classical communication.

\section{Capacities assisted by forward communication}\label{s:quno}

In this section we analyze the role of free forward classical
communication and show that for any $x$, $y$ and $p$,  
\begin{eqnarray}
C(x,y,p) &=& y + C(x,0,p)\label{C1}\;,\\
Q(0,y,0) &=& Q(0,0,0) \label{Q1} \;.
\end{eqnarray}
The first expression states that if the sender is provided with
$y$ classical dits of forward communication per channel use,
the classical capacity of the channel cannot be increased of more than
$y$ dits per channel use.
This result can be interpreted as an 
application of  
the additivity property of the entanglement breaking channels 
\cite{BREAK} to the case
$x,p\neq 0$. 
We provide an explicit proof of Eq.~(\ref{C1}) in Sec.~\ref{s:qunouno}.
Equation~(\ref{Q1}) is a little more subtle: it 
implies that, for $x=p=0$, 
free forward classical communication cannot be used 
to boost the quantum capacity of a channel. This fact was
first pointed out in Ref.~\cite{BENNETT1} and successively in 
Ref.~\cite{BARNUM}
 by explicitly proving that from any 
protocol that uses classical forward
communication one can derive another protocol which does not
use such resource but that achieves, asymptotically, the same
communication rate. In Sec.~\ref{s:qunodue}
 we give an alternative derivation
of this result by direct calculation of the capacity $Q(0,y,0)$.
In Sec.~{\ref{s:disto}} we will prove a stronger version
of this identity by showing that $D(0,y,0)=Q(0,0,0)$.

\subsection{Classical capacity assisted by forward classical 
communication}\label{s:qunouno}
To derive Eq. (\ref{C1}) we notice that the right-hand side
of this equation is a lower bound for the capacity $C(x,y,p)$.
In fact, for each use of channel $\cal M$,  the sender can exploit the forward
communication resource to directly transmit (in average)
$y$ dits.
Moreover, by using the channel $\cal M$ with $x$ dits of feedback and
$p$ e-dits of share entanglement, she/he can still communicate at a rate
$C(x,0,p)$.
Hence to prove the identity in Eq.~(\ref{C1}) we only need to show
that 
\begin{eqnarray}
C(x,y,p) \leqslant y + C(x,0,p)\label{C1uno}\;.
\end{eqnarray}
This can be accomplished, for instance, by considering the capacity 
$C(x,0,p)$ in the absence of free forward communication.
By definition, using the channel $N$ times the sender cannot transmit 
more than $N C(x,0,p)$ dits of classical communication.
Suppose now that she/he decides to use the classical dits transmitted in a
fraction $\gamma\in[0,1]$ of the $N$ channels
 as a resource  to boost the capacity of the
remaining $(1-\gamma)N$ channel uses. The total number of dits
transmitted in the first part of the protocol is $\gamma N C(x,0,p)$,
which provides in average
\begin{eqnarray}
y\equiv\gamma  C(x,0,p)/(1-\gamma)
\label{Ypsilon}
\end{eqnarray}
dits per channel use
of forward communication available as a resource
 for the remaining $(1-\gamma)N$ uses.
In this way, when using these channels
she/he can achieve a capacity  $C(x,y,p)\geqslant
C(x,0,p)$ and hence a total of 
$(1-\gamma) N C(x,y,p)$
dits of classical communication transmitted.
Since this number cannot exceed $N C(x,0,p)$, we have
\begin{eqnarray}
C(x,0,p)\geqslant (1-\gamma) C(x,y,p)\;,
\label{class}
\end{eqnarray}
which yields Eq.~(\ref{C1uno}) 
by solving for $\gamma$ in terms of $y$ through 
Eq.~(\ref{Ypsilon}).

\subsection{Quantum capacity assisted  by forward classical 
communication}\label{s:qunodue}

To prove Eq.~(\ref{Q1}) it
is sufficient to show that the right-hand side term is
greater than or equal to the left-hand side term. In fact
by definition we have that $Q(0,y,0)\geqslant Q(0,0,0)$ 
for all $y$.
We remind the reader that the unassisted capacity $Q=Q(0,0,0)$
of a channel $\cal M$ can be calculated as 
the sup over $n$ successive uses of the channel, 
\begin{eqnarray} Q \equiv \sup_n \; Q_n/n \;,
\label{quantum}
\end{eqnarray}
of the maximum coherent
information~\cite{SETHQ,SCHUMACHER},
\begin{eqnarray} 
Q_n \equiv \max_{\rho\in {\cal H}^{\otimes n}} 
\Big\{ S({\cal  M}^{\otimes n} (\rho))-
S(({\cal M}^{\otimes n}
\otimes \openone_{anc}) (\Phi_\rho))\Big\}\;,
\nonumber\\
\label{q}
\end{eqnarray}
achievable over the set of the
input density matrices  $\rho$ of ${\cal H}^{\otimes n}$.
In Eq.~(\ref{q}) $S(\rho)=-\mbox{Tr}
[\rho  \log_d \rho ]$ is the von Neumann entropy expressed in 
dits, and
$\Phi_\rho$ is a 
purification~\cite{CHUANG}
of $\rho$ defined in the extended 
space obtained by 
adding an ancillary space ${\cal H}_{anc}$ to 
${\cal H}^{\otimes n}$.

An expression analogous to Eq.~(\ref{q}) can be used to compute the 
capacity $Q(0,y,0)$ of the channel $\cal M$ in the presence
of $y$ dits of classical forward communication.
In fact, consider a Hilbert space ${\cal H}^\prime$
of dimension $\Delta\geqslant d^{y}$ and define $\cal T$ the
CPT map  which describes the complete decoherence of the system in
the orthonormal basis $\{ |\omega\rangle\}$
of ${\cal H}^\prime$, i.e.
\begin{eqnarray}
{\cal T}(|\omega\rangle\langle \omega^{\prime}|)\equiv
 \delta_{\omega,\omega^\prime} \; 
|\omega\rangle\langle \omega | \;.
\label{decoherence}
\end{eqnarray}
This map is an entanglement breaking channel~\cite{BREAK}:
when used in the absence of any external resource, it cannot
transfer quantum information but can reliably transmit
(at least) $y$ dits of classical information by encoding them 
into the occupation numbers of the basis $\{ |\omega\rangle\}$.
The super-operator $\overline{\cal M}\equiv{\cal M}\otimes {\cal T}$ 
defines a quantum channel which acts on the input Hilbert 
space ${\cal H}\otimes {\cal H}^\prime$. 
Clearly the quantum capacity $\overline{Q}(0,0,0)$ of this channel
 is at 
least as big as the capacity $Q(0,y,0)$
of the original channel $\cal M$: by employing
$\cal T$ to transmit $\log_d \Delta$ dits of classical
information at each use of $\overline{\cal M}$, we can simulate
the performance of the  channel $\cal M$ when it is 
assisted by $y$ dits of
forward communication.
The identity~(\ref{Q1}) can be hence proved 
by showing that 
\begin{eqnarray}
\overline{Q}(0,0,0) \leqslant {Q}(0,0,0)
\label{prova}\;.
\end{eqnarray}
In the following we will do that by expressing $\overline{Q}(0,0,0)$
in terms of the coherent information
of $\overline{\cal M}$ as in Eq.~(\ref{q}). 

 Given the $n$-elements vector 
$\vec{\omega}\equiv (\omega_1,\cdots,\omega_n)$, define
 $|\vec{\omega}\rangle\equiv \otimes_{j=1}^n|\omega_j\rangle$ the
orthonormal basis of ${{\cal H}^{\prime}}^{\otimes n}$ obtained by
taking $n$ copies of the basis $\{|\omega_j\rangle \}$ 
of ${\cal H}^\prime$.
The map $\overline{\cal M}^{\otimes n}$ transforms any
density matrix $R$ of $({\cal H} \otimes {\cal H}^\prime)^{\otimes n}$ 
according to
\begin{eqnarray} 
\overline{\cal M}^{\otimes n}(R) \equiv  \sum_{\vec{\omega}} 
\lambda_{\vec{\omega}} 
{\cal M}^{\otimes n}(\rho_{\vec{\omega}}) 
\otimes |\vec{\omega}\rangle \langle \vec{\omega}|
\label{overmap}\;,
\end{eqnarray}
where 
\begin{eqnarray}
\lambda_{\vec{\omega}} \,\rho_{\vec{\omega}} 
\equiv \langle \vec{\omega} | R | \vec{\omega} \rangle
\;,\label{Revoluto}
\end{eqnarray}
is the unnormalized density matrix of ${\cal H}^{\otimes n}$ 
obtained by projecting $R$ into $|\vec{\omega}\rangle$, 
$\lambda_{\vec{\omega}}$ being the probability associated with
such a projection. From the orthogonality of $\{|\vec{\omega}\rangle\}$
we can thus express the von Neumann entropy  of 
$\overline{\cal M}^{\otimes n}(R)$ as~\cite{CHUANG}
\begin{eqnarray} 
S(\overline{\cal M}^{\otimes n}(R)) = 
H(\lambda_{\vec{\omega}})+ \sum_{\vec{\omega}} \lambda_{\vec{\omega}}
S({\cal M}^{\otimes n}(\rho_{\vec{\omega}}))\;,
\label{output}
\end{eqnarray}
with 
\begin{eqnarray} 
H(\lambda_{\vec{\omega}}) \equiv - \sum_{\vec{\omega}} 
\lambda_{\vec{\omega}}
\log_d  \lambda_{\vec{\omega}} \;,
\label{shannon}
\end{eqnarray}
the Shannon entropy associated to the probabilities 
$\lambda_{\vec{\omega}}$.

Consider now a purification $\Phi_R\equiv|\Phi_R\rangle \langle \Phi_R|$ 
of $R$. Its
projection into the state $|\vec{\omega}\rangle$ of 
${{\cal H}^{\prime}}^{\otimes n}$ gives 
\begin{eqnarray} 
\langle \vec{\omega} |\Phi_R\rangle\equiv \sqrt{\lambda_{\vec{\omega}}} \; |
\Phi_{\rho_{\vec{\omega}}}\rangle
 \;, \label{purificazione1}
\end{eqnarray}
where $|\Phi_{\rho_{\vec{\omega}}}\rangle \in 
{\cal H}^{\otimes n}\otimes {\cal H}_{anc}$ is a purification of
the density matrix $\rho_{\vec{\omega}}$ of Eq.~(\ref{Revoluto}).
From Eq.~(\ref{overmap}) derives thus
\begin{eqnarray} 
(\overline{\cal M}^{\otimes n}\otimes \openone_{anc})
(\Phi_R) = \sum_{\vec{\omega}} \lambda_{\vec{\omega}} 
{\cal M}^{\otimes n}(\Phi_{\rho_{\vec{\omega}}})
\otimes |\vec{\omega}\rangle\langle \vec{\omega} |
 \;, \label{evoluto}
\end{eqnarray}
with  $\Phi_{\rho_{\vec{\omega}}}\equiv 
|\Phi_{\rho_{\vec{\omega}}}\rangle \langle 
\Phi_{\rho_{\vec{\omega}}}|$, and hence
\begin{eqnarray} 
&&S((\overline{\cal M}^{\otimes n}\otimes \openone_{anc})
(\Phi_R)) \label{output1} \\
&&\qquad =H(\lambda_{\vec{\omega}}) 
+\sum_{\vec{\omega}} \lambda_{\vec{\omega}}
S(({\cal M}^{\otimes n}\otimes \openone_{anc})
(\Phi_{\rho_{\vec{\omega}}}))\;.\nonumber
\end{eqnarray}
From Eqs.~(\ref{output}) and (\ref{output1}) we finally obtain
\begin{eqnarray} 
&&S((\overline{\cal M}^{\otimes n})(R))
-S((\overline{\cal M}^{\otimes n}\otimes \openone_{anc})
(\Phi_R)) \label{coherent} \\
&&\quad =
\sum_{\vec{\omega}} \lambda_{\vec{\omega}}
\left[ S({\cal M}^{\otimes n}
\rho_{\vec{\omega}}) - 
S(({\cal M}^{\otimes n}\otimes \openone_{anc})
(\Phi_{\rho_{\vec{\omega}}})) \right] 
\nonumber \;,
\end{eqnarray}
which  shows that the coherent information of the map
$\overline{\cal M}^{\otimes n}$ relative to the
input state  $R$ of  $({\cal H} \otimes {\cal H}^\prime)^{\otimes n}$  
can be expressed as a convex combination of the 
coherent informations of the map ${\cal M}^{\otimes n}$. 
According to~\cite{SETHQ} 
the quantum capacity $\overline{Q}(0,0,0)$
of $\overline{\cal M}$
is obtained by maximizing over $R$ the left-hand side term 
of Eq.~(\ref{coherent}) and then by taking the sup over $n$.
The inequality~(\ref{prova}) finally derives by
noticing that $Q_n$ of 
Eq.~(\ref{q})  is greater than or equal to 
 the right-hand side of
 Eq.~(\ref{coherent}) for all $R$ and for all $n$ integer.

\section{Asymptotic limits and trade-off curves for
capacities}\label{s:tradeoff}

In this section we analyze in detail the relations between the
capacities associated with different resources. We will focus
mostly on the properties of $Q(x,y,p)$. For the sake of
simplicity part of the material relative to the no-feedback case
($x=0$) 
has been postponed in Appendix~\ref{s:appfinale}.

\subsection{Quantum capacity} \label{s:quantumresp}
\begin{figure}[t]
\begin{center}
\epsfxsize=.7\hsize\leavevmode\epsffile{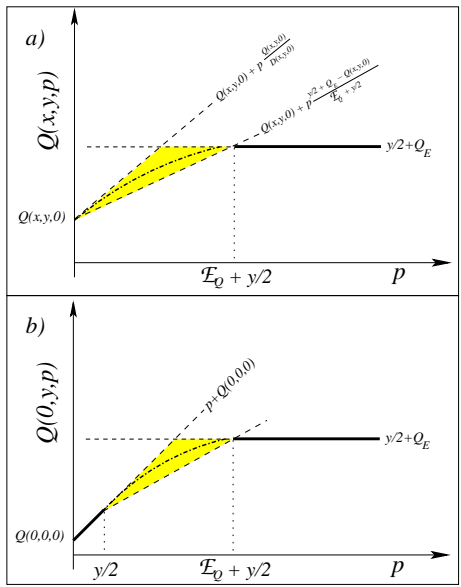}
\end{center} 
\caption{Plot of  
the quantum capacity $Q(x,y,p)$ of the channel $\cal M$
as a function of the shared entanglement resource parameter $p$.
Plot $a)$: according to Eq.~(\ref{Q8}),
for  $p\geqslant y/2 + {\cal E}_Q$ the function
$Q(x,y,p)$ achieves the asymptotic value  $y/2 + Q_E$.
For  $p\leqslant y/2 + {\cal E}_Q$, $Q(x,y,p)$
lies in the gray region defined by the upper bounds of Eqs.~(\ref{Q2})
and (\ref{Q3}) and by the lower bound of Eq.~(\ref{Q10}). 
The dot-dashed line is a tentative plot of $Q(x,y,p)$ which
takes into account these bounds and the concavity with respect to $p$.   
Plot $b)$: no-feedback case ($x=0$).
Here we can use Eq.~(\ref{Q77}) to determine 
the behavior of $Q(0,y,p)$ for small $p$. The lower bound for 
$p\in[y/2,y/2+{\cal E}_Q]$ derives from the concavity of $Q(0,y,p)$.
The plots units are arbitrary.}
\label{f:fig2}
\end{figure}

The behavior of $Q(x,y,p)$ as a function of the parameter $p$ 
is sketched in Fig.~\ref{f:fig2}. 
We begin by showing that for all $x$, $y$ and
$p$ one has
\begin{eqnarray}
p \frac{Q(x,y,0)}{D(x,y,0)} + Q(x,y,0)
\geqslant  Q(x,y,p) \label{Q2}\;,
\end{eqnarray}
where $D(x,y,0)$ is the distillability of $\cal M$ defined in Eq.~(\ref{distillato}).
This relation essentially states that $p$ e-dits  cannot produce more than
$p$  qudits of quantum information. To prove it we proceed analogously to
Sec.~\ref{s:qunouno} and consider $N>>1$ uses of the channel $\cal M$ in the
presence of $x$ dits of feedback, $y$ dits of classical forward information 
and
no shared entanglement. By definition the sender cannot 
transfer more than $N Q(x,y,0)$ qudits. Suppose now that she/he decides to
use a fraction $\gamma$ of the $N$ channels to share with the receiver some
maximally entangled state that will then be employed as a resource for the
remaining $(1-\gamma)N$ uses. 
Since the total number of e-dits transmitted
on $\gamma N$ channels is at most   
$\gamma 
N D(x,y,0)$, they obtain on average
\begin{eqnarray}
p\equiv \gamma D(x,y,0) /(1-\gamma)
\label{pshared}
\end{eqnarray}
e-dits per channel as a resource for the $(1-\gamma)N$ remaining channels.
In the second part of the protocol the communicating parties can thus achieve a quantum 
capacity $Q(x,y,p)$ which is greater than the initial $Q(x,y,0)$.
The total number of qubits transmitted in this way is hence equal
to $(1-\gamma)N Q(x,y,p)$. Equation~(\ref{Q2}) then follows by requiring this
quantity to be smaller than the maximum number of qudits transmittable
, i.e. $N Q(x,y,0)$.

The upper bound of Eq.~(\ref{Q2}) is not always 
tight and for large values of $p$
is replaced by
\begin{eqnarray}
y/2 + Q_E &\geqslant& Q(x,y,p)\label{Q3} \;.
\end{eqnarray}
To prove this inequality
 consider the scenario where the communicating parties
are provided with $x$ dits of feedback, $y=0$ 
forward classical communication,
and $p$ e-dits of shared entanglement. In this case, 
by using the channel $N$ times
the sender can transfer
at most $N Q(x,0,p)$ qudits. Compare this with the number of
qubits that can be transmitted when a fraction of the
channels are employed to produce some dits
of classical communication as a resource for the remaining
channels.
In the limit $N>>1$ we get
\begin{eqnarray}
y\frac{Q(x,0,p)}{C(x,0,p)} + Q(x,0,p) &\geqslant& Q(x,y,p)\label{Q4} \;.
\end{eqnarray}
which for $p\rightarrow\infty$ becomes 
(see Eqs.~(\ref{utile}) and (\ref{bow1})) 
\begin{eqnarray}
y/2 + Q_E &\geqslant& Q(x,y,\infty)\label{Q5} \;.
\end{eqnarray}
The inequality~(\ref{Q3}) derives now from the monotonicity of $Q(x,y,p)$
with respect to $p$.

Simple lower bounds for $Q(x,y,p)$ are obtained by exploiting 
teleportation.
In fact if $p \geqslant y/2$, the sender can use
the $y$ dits of forward communication and $y/2$ e-dits of
shared entanglement to teleport $y/2$ qudits to the receiver.
At this point she/he can still use the channel $\cal M$ to transmit
at a rate equal to $Q(x,0,p-y/2)$. i.e.
\begin{eqnarray}
Q(x,y,p)\geqslant y/2 + Q(x,0,p-y/2) \quad \mbox{for } p \geqslant y/2 
\label{Q6} \;.
\end{eqnarray}
Analogously one can show that 
\begin{eqnarray}
Q(x,y,p)\geqslant p + Q(x,y-2p,0) \quad \mbox{for } p \leqslant y/2 
\label{Q7} \;.
\end{eqnarray}

Consider now the case $p\geqslant y/2 + {\cal E}_Q$, with ${\cal E}_Q$ 
defined as in Sec.~\ref{s:def}. In this limit, Eq.~(\ref{bow1}) implies 
$Q(x,0,p-y/2) = Q_E$, and by 
confronting Eq.~(\ref{Q3}) with Eq.~(\ref{Q6}) we
obtain 
\begin{eqnarray}
Q(x,y,p)= y/2 + Q_E \quad \mbox{for } 
p\geqslant y/2 + {\cal E}_Q \label{Q8} \;.
\end{eqnarray}
The right-hand side of this expression is thus an 
asymptote for the capacity 
(see Fig.~\ref{f:fig2}). It is achieved for a critical 
value of $p$ which is smaller than $y/2 + {\cal E}_Q$
and greater than the intercept between
the asymptote and the upper bound of Eq.~(\ref{Q2}) 
[Notice that by imposing  $y/2 + {\cal E}_Q$ to be greater than this
point one gets
${\cal E}_Q \geqslant Q_E -Q(x,y,0)$ which
 gives Eq.~(\ref{bowenin1}) for $x=y=0$].
From these considerations and from the concavity of $Q(x,y,p)$, 
we can  now 
establish the linear lower bound of plot $a)$ of Fig.~\ref{f:fig2},
i.e.
\begin{eqnarray}
 Q(x,y,p) \geqslant p \; \frac{y/2 + Q_E -Q(x,y,0)}{y/2+{\cal E}_Q} 
+ Q(x,y,0)
 \label{Q10} \;,
\end{eqnarray}
for $p\leqslant y/2 + {\cal E}_Q$.
For $x=0$ this inequality can be improved by means of 
Eqs.~(\ref{bowenin1}) and (\ref{Q1}). In fact in this limit
Eq.~(\ref{Q7}) yields 
\begin{eqnarray}
Q(0,y,p)\geqslant p + Q(0,y-2p,0) =  p + Q(0,0,0) 
\label{Q77} \;,
\end{eqnarray}
for $p \leqslant y/2$, which by comparison with Eqs.~(\ref{Q2}) 
and (\ref{distillato1}) implies 
\begin{eqnarray}
Q(0,y,p)= p + Q(0,0,0) \qquad \mbox{for } p \leqslant y/2
\label{Q777} \;.
\end{eqnarray}
This equation
generalizes the identity~(\ref{Q1}) 
to the case of shared entanglement.
In Appendix~\ref{s:appfinale} we provide a more detailed 
analysis of the no-feedback case, by analyzing 
a conjecture proposed by Bowen~\cite{BOWEN1}.

\subsubsection{Quantum capacity with forward classical communication}
\label{s:quantumresy}
\begin{figure}[t]
\begin{center}
\epsfxsize=.7 \hsize\leavevmode\epsffile{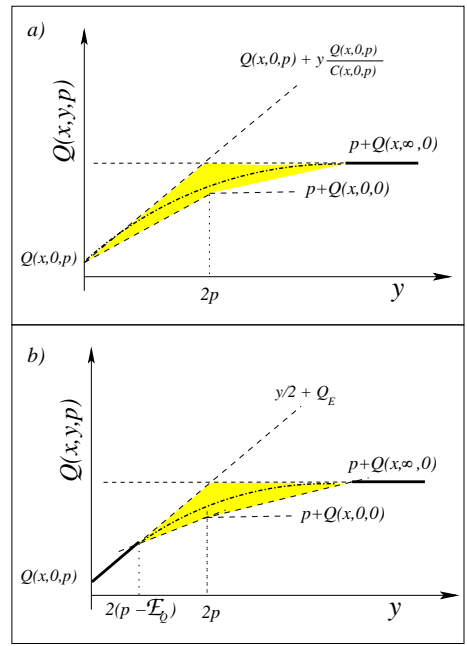}
\end{center} 
\caption{Plot of  
the quantum capacity $Q(x,y,p)$ of the channel $\cal M$
as a function of the forward communication resource parameter $y$.
This quantity is increasing, concave and lies in the gray region.
Plot $a)$: 
according to Eq.~(\ref{YQ}) in the limit $y>>2p$, $Q(x,y,p)$ reaches the
asymptotic value $p+Q(x,\infty,0)$. 
Plot $b)$: for $p\geqslant {\cal E}_Q$
and $y\leqslant 2(p-{\cal E}_Q)$ the value of $Q(x,y,p)$ is determined
by Eq.~(\ref{Q8}). Units and symbols are defined as in
 Fig.~\ref{f:fig1}.}
\label{f:fig4}
\end{figure}

The dependence of $Q(x,y,p)$ with respect to the resource $y$ has been
plotted in Fig.~\ref{f:fig4}. The asymptote
\begin{eqnarray}
 Q(x,\infty,p) = \; p + Q(x,\infty,0)
 \label{YQ} \;,
\end{eqnarray}
is derived by considering Eqs.~(\ref{Q2}) and (\ref{Q7}) in the limit 
$y>>2p$.
For $x\neq0$, we do not have a method to characterize the critical 
$y$ for which this asymptotic regime is achieved. However, 
Eq.~(\ref{Q777})
shows that in the case of no-feedback ($x=0$) this quantity is smaller
than $2p$. 
The linear upper bound given in the plot $a)$ of Fig.~\ref{f:fig4}
is provided by Eq.~(\ref{Q4}).
For $p\geqslant {\cal E}_Q$ 
and $y \leqslant(p-{\cal E}_Q)/2$ the 
value of $Q(x,y,p)$ is determined by Eq.~(\ref{Q8}) while
Eq.~(\ref{Q3}) gives a better lower bound for $Q(x,y,p)$
than~(\ref{YQ}) (see plot $b)$ of Fig.~\ref{f:fig4}).

\subsection{Classical capacity} \label{s:classresp}

Equation~(\ref{C1}) allows us to focus only on the  case $y=0$.
A lower bound for $C(x,0,p)$ derives from the concavity with respect to
$p$ and from the
definition of ${\cal E}_C$ given in Eq.~(\ref{bow1}),
i.e.
\begin{eqnarray}
{C(x,0,p)}\geqslant p \frac{C_E -C}{{\cal E}_C} + C
\label{lower}\;.
\end{eqnarray}
As in the case of Eq.~(\ref{Q4}) we can
establish the following inequality,
\begin{eqnarray}
p\frac{C(x,0,0)}{D(x,0,0)} + C(x,0,0) &\geqslant& C(x,0,p)\label{CCC} \;,
\end{eqnarray}
with $D(x,0,0)$ the distillability of the channel $\cal M$ achieved
by using on average $x$ dits of classical feedback.
Equation~(\ref{CCC}) can be derived by comparing
 the capacity $C(x,0,0)$ with the number of
bits transmittable  with a protocol
where the sender and receiver employ part of the channel uses 
to share maximally entangled pairs.

\section{Identities for distillability}\label{s:disto}
From the definition of $D(x,y,p)$ of Eq.~(\ref{distillato}) one
can prove that for any $x$, $y$ and $p$ the following identities
hold
\begin{eqnarray}
&&D(x,\infty,p) = Q(x,\infty,p)
\label{D10}\;, \\
&&D(x,y,p) = p+ D(x,y,0)\;,  \label{D11}\\
&&D(0,y,0) = D(0,0,0)=Q(0,0,0)\;.
\label{D12}
\end{eqnarray}
The first identity is a trivial generalization of Eq.~(\ref{qudue}).
It derives by noticing that, with infinite free forward classical
communication, each  of the maximally entangled state
distilled from the channel can be used
to teleport one qudit of quantum information.
This implies that $Q(x,\infty,p)\geqslant D(x,\infty,p)$ which together
with Eq.~(\ref{distillato1}) gives
the relation~(\ref{D10}). 
The identity~(\ref{D11}) is the analog of Eq.~(\ref{C1}) in the
context of the distillability of a channel. It states that
adding $p$ e-dits  of shared entanglement per channel use 
to the resources, 
the distillability of $\cal M$  cannot be increased by more than
$p$ e-dits per channel use. 
Since the proof of this identity 
can be obtained,
{\em mutatis mutandis}, from the proof of Eq.~(\ref{C1}) we will skip it.

A less trivial identity is Eq.~(\ref{D12}) which can be seen as
a stronger version of Eq.~(\ref{Q1}). For the special case of 
generalized depolarizing channels it was first proved in Ref.~\cite{DIST}.
It implies that free forward communication is not sufficient to
extract maximally entangled states at a rate higher than the
unassisted capacity of a channel.
To prove it, consider the inequality~(\ref{Q2}) for
 $x=0$ and $p\leqslant y/2$. Using the properties~(\ref{Q1}) and
(\ref{Q777}) we get
\begin{eqnarray}
{Q(0,0,0)}\geqslant {D(0,y,0)}
\label{Q222}\;,
\end{eqnarray}
which, according to Eq.~(\ref{distillato1}) gives the identity~(\ref{D12}). 

Equations~(\ref{D11}) shows that for $p >>{\cal E}_Q, {\cal E}_C$ 
and finite values
of $y$ the distillability $D(x,y,p)$ is  
bigger than the corresponding capacities
$Q(x,y,p)$ and $C(x,y,p)$ (these two quantities saturate respectively to
$y/2 + Q_E$ and $y+ C_E$). 
However, according to this same equation, $p$ cannot
be considered as a proper resource for distillation protocols.
The interesting cases are hence those where $p=0$.
Here Eqs.~(\ref{D10}) and (\ref{D12})
show that there are scenarios where the distillability 
coincides with the quantum capacity of the channel and can be thus strictly
lower than the classical capacity.

\section{Conclusion}\label{s:con}
In this paper we studied the performance of a quantum channel $\cal M$ in the
presence of external resources.  In particular we focused on the dependence of
its capacities with respect to the resource parameters, deriving some
inequalities and trade-offs. We have also introduced the concept of
distillability of a quantum channel, by maximizing the distillable
entanglement one can get at the input and output ports of $\cal M$ 
over all purification protocols that exploit only finite amount of
resource per channel use.

\appendix
\section{Distillability}\label{s:app2}
A purification protocol $\cal P$ acting on $k$ copies of 
 $R^\prime$ produces
$m(k)$ copies of a given maximally entangled
state $|\Psi\rangle$ of ${\cal H}\otimes {\cal H}$ with 
fidelity $F(k)$ which approaches unity in the limit of large
$k$~\cite{DIST}. 
Here we are interested only on those protocols $\cal P$ that 
operate on  ${R^\prime}^{\otimes k}$
by employing in total $kn x$ dits of classical
feedback, $kn y$ dits of free classical forward communication,
and $kn p$ e-dits of shared entanglement.
The quantity $P_{x,y,p}\left[R^\prime\right]$ is then defined by
the ratio $m(k)/k$ by  
optimizing  $m(k)$  over all $\cal P$ and 
by considering the limit $k\rightarrow \infty$.
Thus besides the sup over $n$ of Eq.~(\ref{distillato}), the computation of 
$D(x,y,p)$ also
requires a regularization over the parameter $k$ defined above.

\section{Concavity}\label{s:app1}
In this Appendix we show the joint concavity of
$Q(x,y,p)$,  $C(x,y,p)$ and $D(x,y,p)$. 

For the sake of simplicity write 
$Q(x_i)\equiv Q(x,y,p)$ where for $i=1,2,3$, $x_1\equiv x$,
$x_2\equiv y$ and $x_3\equiv p$.
Consider the case where the two communicating parties have access to 
$N>>1$ uses of the channel $\cal M$ 
and hence to 
$X_N(i)\equiv N x_i$ units of the $i$th 
resource.
By definition, the sender cannot transfer more than
$NQ(x_i)$ qudits to the receiver. Suppose now that they divide
 the set of
$N$ channels into
two groups: the group $A$ with  $N_A$ channels and the group $B$ 
with $N_B\equiv N-N_A$ channels. 
Moreover, when operating the channels of $A$, 
the sender and receiver decide
to employ only $X_A(i) \leqslant X_N(i)$ 
units of the $i$th resource, corresponding to an
 average of $x_i^A\equiv X_A(i)/N_A$ for this set.
The remaining $X_B(i) \equiv X_N(i)-X_A(i)$ units of the $i$th resource 
are instead used when operating the channels of $B$ 
(which gives an average
of $x_i^B\equiv X_B(i)/N_B$ units per channel use for this set).
In the limit of very large $N$ the quantum capacity associated with
the channels of the set $A$ is thus given  by
$Q(x_i^A)$: the maximum number of qudits
that can be transmitted using these channels is thus
$N_A Q(x_i^A)$. Analogously for the set $B$ we have
a maximum number of $N_BQ(x_i^B)$ qudits transmitted.
The sum of these two quantities cannot exceed the
optimal value  $NQ(x_i)$
and we obtain the inequality
\begin{eqnarray}
Q(x_i) \geqslant \frac{N_A}{N} Q(x_i^A)
+ \frac{N_B}{N} Q(x_i^B)\;,
\label{qqq}
\end{eqnarray}
which, since $x_i=x_i^A N_A/N + x_i^B N_B/N$,
proves the joint concavity of $Q(x_i)$.
The same procedure can be applied 
in the case of $C(x,y,p)$ and $D(x,y,p)$.
Notice that joint concavity with respect to $x$, $y$ and $p$ 
implies concavity in each of these variables 
(see for instance~\cite{CHUANG}).

\section{No-feedback case and Bowen conjecture}\label{s:appfinale}

In this Appendix we discuss a conjecture proposed in~\cite{BOWEN1}
showing that in the no-feedback
regime ($x=0$) it allows one to solve exactly the value of $Q(0,y,p)$.

The Bowen conjecture implies that for any channel $\cal M$ 
the value of ${\cal E}_Q$ of 
Eq.~(\ref{bowenin2}) is given by
\begin{eqnarray}
{\cal E}_Q = Q_E -Q \;,
\label{CONJ}
\end{eqnarray}
where $Q_E$ and $Q$ are, respectively, the entanglement assisted 
capacity
$Q(0,0,\infty)$ and the unassisted capacity $Q(0,0,0)$ of the channel.
On the one hand the validity of this conjecture was challenged recently by the
results of~\cite{DHW} which seem to indicate that this relation 
does not hold for generic $\cal M$.
On the other hand  we know that there are examples of channels
(e.g.  dephasing and erasure channels~\cite{BOWEN1,ERASURE})
which satisfy Eq.~(\ref{CONJ}). In any case, whether or not the Bowen conjecture
is a true statement for all CPT map $\cal M$,
it is worth studying its consequences on the quantum capacity.

If Eq.~(\ref{CONJ}) is true  we can use the inequalities derived in 
Sec.~(\ref{s:tradeoff}) to verify that the following identity applies
\begin{eqnarray}
Q(0,0,p)= p + Q(0,0,0) \quad \mbox{for } p\leqslant {\cal E}_Q\;,
\label{CONJ1}
\end{eqnarray}
(this follows for instance by noticing that, for $x=0$,
 Eq.~(\ref{CONJ}) implies that the gray region in the plot $b)$ of 
Fig.~(\ref{f:fig2}) 
vanishes). On the other hand, one can verify that if Eq.~(\ref{CONJ1}) holds,
then Eq.~(\ref{CONJ}) follows. In other words, the Bowen conjecture~(\ref{CONJ}) 
is equivalent to the property~(\ref{CONJ1}).
Moreover, by replacing Eqs.~(\ref{Q1}) and (\ref{CONJ}) in Eq.~(\ref{Q10}) 
we obtain 
\begin{eqnarray}
 Q(0,y,p) \geqslant p  + Q(0,0,0)
 \label{Q100} \;,
\end{eqnarray}
for all $p\leqslant y/2+{\cal E}_Q$. As a matter of fact, the inequality
in Eq.~(\ref{Q100}) can be replaced with an identity by observing that,
according to Eqs.~(\ref{Q2}) and (\ref{D12})
the right hand side of this expression is also an upper bound for
$Q(0,y,p)$.

\begin{figure}[t]
\begin{center}
\epsfxsize=.7\hsize\leavevmode\epsffile{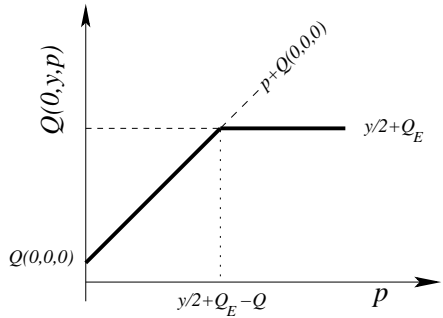}
\end{center}
\caption{Plot of $Q(0,y,p)$ under the 
conjecture of Eq.~(\ref{CONJ}). 
In this case the capacity  
is determined by the concavity and the upper bound~(\ref{Q2}): 
its value is given by Eq.~(\ref{Q101}). 
Compare this plot with plot $b)$ of Fig.~\ref{f:fig2} where the 
conjecture~(\ref{CONJ}) was not taken into account.}
\label{f:fig5}
\end{figure}

The above results show that the conjecture~(\ref{CONJ}) is
equivalent to the property  (see Fig.~\ref{f:fig5})
\begin{eqnarray}
 Q(0,y,p) =\left\{\begin{array}{cl} p  + Q(0,0,0)& \mbox{ for } 
p\leqslant y/2+Q_E-Q \\ \\
y/2 + Q_E  & \mbox{ for } 
p\geqslant y/2+Q_E-Q \;, 
\end{array}\right.
\label{Q101} 
\end{eqnarray}
which implies that 
free forward communication does not 
increase the quantum 
capacity of the channel also in the presence of
prior share entanglement (apart from a trivial contribution due to
direct teleportation of the entanglement resource in excess).
On one hand, for $y>0$, Eq.~(\ref{Q777}) shows that
Eq.~(\ref{Q101}) is verified at least for $p\leqslant y/2$. 
On the other hand, for  $y=0$ and $C(0,0,0)\neq 0$, 
one can show that Eq.~(\ref{Q4}) implies that Eq.~(\ref{Q101}) applies
at least for $p$ sufficiently small.

\acknowledgments
The author would like to thank  R. Fazio, A. Harrow, S. Mancini, H. Nakano and
D. Vitali for 
comments and discussions.
This work was supported by the European 
Community under contracts IST-SQUIBIT,
IST-SQUBIT2, and RTN-Nanoscale Dynamics.


\begin{references}
\bibitem{CHUANG} M. A. Nielsen and I. L. Chuang, {\em Quantum
Computation and Quantum Information} (Cambridge University Press,
Cambridge, 2000).
\bibitem{SHOR} C. H. Bennett and P. W. Shor, 
IEEE Trans. Inf. Theory
{\bf 44}, 2724 (1998).
\bibitem{TELE}C. H. Bennett,
G. Brassard, C. Cr\'epeau, R. Jozsa, A. Peres, and W.
K. Wootters, Phys. Rev. Lett. {\bf 70}, 1895 (1993).
\bibitem{SUP}C. H. Bennett and 
S. J. Wiesner, Phys. Rev. Lett. {\bf 69}, 2881 (1992).
\bibitem{BENNETT1}
C. H. Bennett, P. W. Shor, J. A. Smolin,
and A. V. Thapliyal, Phys. Rev. Lett. {\bf 83}, 3081 (1999);
IEEE Trans. Inf. Theory {\bf 48}, 2637 (2002).
\bibitem{SHORCE}P. W. Shor,  ``
The classical capacity achievable by a quantum channel assisted by limited entanglement''
in O. Hirota, ed., {\em Quantum Information, Statistics, Probability}
(Rinton Press, Princeton, 2004).
\bibitem{DIST} C. H. Bennett, 
D. P. DiVincenzo, J. A. Smolin, and W. K. Wootters, 
Phys. Rev. A {\bf 54}, 3824 (1996).
\bibitem{DIST1}
C. H. Bennett, G. Brassard, S. Popescu, B. Schumacher,
J. A. Smolin, and W. K. Wootters, Phys. Rev. Lett. {\bf 76} 722 (1996).
P. Horodecki and R. Horodecki, Q. Inf. and Comp. {\bf 1}, 45 (2001).
\bibitem{BOWEN}G. Bowen, IEEE Trans. Inf. Theory {\bf 50}, 2429 (2004); 
eprint 
quant-ph/0305176.
\bibitem{DEVETAK}I. Devetak and P. W. Shor, 
eprint quant-ph/0311131.
\bibitem{DHW}I. Devetak, A. W. Harrow, and A. Winter, eprint quant-ph/0308044.
\bibitem{HARROW}A. Harrow, Phys. Rev. Lett. {\bf 92}, 097902 (2004); 
A. Harrow,
P. Hayden, and D. Leung, Phys. Rev. Lett. {\bf 92}, 187901 (2004);
D. Kretschmann and
R. F. Werner, New J. Phys. {\bf 6} 26 (2004).
\bibitem{BENNEW}C. H. Bennett, I. Devetak, P. W. Shor, and J. A. Smolin, 
eprint quant-ph/0406086.
\bibitem{SETHQ} S. Lloyd, Phys. Rev. A {\bf 55}, 1613 (1997);
H. Barnum, M. A. Nielsen, and B. Schumacher, 
Phys. Rev. A {\bf 57}, 4153
(1998); P. W. Shor, available at http://
www.msri.org/publications/ln/msri/2002/
quantumcrypto/shor/1/;
I. Devetak, to appear in IEEE Trans. Inf. Theory, eprint quant-ph/0304127.
\bibitem{HSW}A. S. Holevo, 
IEEE Trans. Inf. Theory {\bf 44}, 269
  (1998); P. Hausladen,
 R. Jozsa, B. Schumacher, M. Westmoreland
and W. K. Wootters, Phys. Rev. A {\bf 54}, 1869 (1996); 
B. Schumacher and M. D. Westmoreland, Phys. Rev. A {\bf 56},
131 (1989).
\bibitem{ADAMI} C. Adami and N. J. Cerf, Phys. Rev. A {\bf 56} 3470 (1997). 
\bibitem{BOWEN1}G. Bowen, Phys. Rev. A {\bf 66} 052313 (2002). 
\bibitem{BREAK}P. W. Shor, J. Math. Phys. {\bf 43}, 4334 (2002); 
M. Horodecki, P. W. Shor, and M. B. Ruskai,
Rev. Math. Phys. {\bf 15}, 629 (2003).
\bibitem{BARNUM}H. Barnum, E. Knill, and M. A. Nielsen, IEEE
Trans. Inf. Theory {\bf 46}, 1317 (2000).
\bibitem{SCHUMACHER}B. Schumacher 
and M. A. Nielsen, Phys. Rev. A
{\bf 54}, 2629 (1996).
\bibitem{ERASURE}C. H. Bennett, D. P. DiVincenzo, and J. A.
 Smolin, Phys. Rev. Lett. {\bf 78} 3217 (1997).
\end{references}
\end{document}